\documentclass[12pt,preprint]{aastex}
\shorttitle{CARMA OBSERVATIONS OF ORION-KL}
\shortauthors{Friedel \& Snyder}
\begin{document}
\newcommand\acetone{(CH$_3$)$_2$CO}
\newcommand\dme{(CH$_3$)$_2$O}
\newcommand\mef{HCOOCH$_3$}
\newcommand\fa{HCOOH}
\newcommand\mtoh{CH$_3$OH}
\newcommand\kms{km s$^{-1}$}
\newcommand\jbm{Jy bm$^{-1}$}
\newcommand\vycn{C$_2$H$_3$CN}
\newcommand\etcn{C$_2$H$_5$CN}
\newcommand\etoh{C$_2$H$_5$OH}
\newcommand\acal{CH$_3$CHO}
\newcommand\vlsr{$v_{\rm LSR}$}

\title{HIGH RESOLUTION $\lambda$=1MM CARMA OBSERVATIONS OF LARGE MOLECULES IN ORION-KL}

\author{D. N. Friedel\altaffilmark{1} and L. E. Snyder\altaffilmark{1}}

\altaffiltext{1}{Department of Astronomy, 1002 W. Green St., University of
Illinois, Urbana IL 61801\\
email: friedel@astro.uiuc.edu, snyder@astro.uiuc.edu}

\begin{abstract}
We present high resolution, Combined Array for Research in Millimeter-Wave Astronomy (CARMA), $\lambda$=1mm observations of several molecular species toward Orion-KL. These are the highest spatial and spectral resolution 1mm observations of these molecules to date. Our observations show that ethyl cyanide [\etcn] and vinyl cyanide [\vycn] originate from multiple cores near the Orion hot core and IRc7. Additionally we show that dimethyl ether [\dme] and methyl formate [\mef] originate from IRc5 and IRc6 and that acetone [\acetone] originates only from areas where both N-bearing and O-bearing species are present.

\end{abstract}


\section{INTRODUCTION}
The Orion-KL region is the closest ($\sim$480 pc) site of massive star formation to us \citep{genzel81}. There are several cloud components (e.g. hot core, compact ridge, extended ridge, and plateau) which are associated with Orion-KL, each with different chemical and physical properties \citep[e.g.][]{blake87}. The two most chemically interesting components, the hot core and compact ridge, are separated by only $\sim$5800 AU. Large oxygen bearing species (e.g.\ methyl formate [\mef] and dimethyl ether [\dme]) have been observed primarily toward the compact ridge \citet[e.g.][]{liu02}, while large nitrogen bearing species (e.g.\ ethyl cyanide [\etcn]) are located toward the hot core \citep[e.g.][]{blake87}.

The chemical pathways to most large detected interstellar molecules in hot molecular cores are not known or are poorly understood \citep[e.g][]{quan07}. The spatial location of a molecular species, in relation to continuum, shocks, and other species, can give indications of its formation mechanism. In order to further our understanding of these formation mechanisms we have conducted high resolution observations of several large molecular species: acetone [\acetone], dimethyl ether [\dme], methyl formate [\mef], ethyl cyanide [\etcn], methanol [\mtoh], formic acid [\fa], acetaldehyde [\acal], and ethanol [\etoh].

While many of these molecular species have been observed at moderate resolution \citep[e.g.][]{liu02,blake96} \acetone\ has not. Its original detection in this source by \citet{friedel05} was with a rather large beam ($\sim12\arcsec\times9\arcsec$) and indicated that its emission was not associated with either the hot core or compact ridge. In this paper we present the highest spectral resolution 1mm observations of these molecular species to date.

\section{OBSERVATIONS}
The observations were taken in 2007 March with the Combined Array for Research in Millimeter-Wave Astronomy (CARMA). All data were acquired in two $\sim$5 hour long tracks (one for \dme, \etcn, \acetone, and continuum and one for \mef, \mtoh, \acal, \etoh, and \fa) with 9 antennas and a phase center of $\alpha$(J2000) = $05^h35^m14^s.35$ and $\delta$(J2000) = $-05{\degr}22{\arcmin}35{\arcsec}.0$. The typical synthesized beam was $\sim2.5\arcsec\times0.85\arcsec$ (54 times higher resolution than the original acetone detection). The $u-v$ coverage of the observations gives projected baselines of 5.8-196.5 k$\lambda$ (7.5-255.5 m). Thus any structure larger than $\sim16\arcsec$ will be resolved out by the array. We compared our flux values to those of single dish observations of \citet{sutton87} and determined that we are resolving out little flux ($<15$\%) from most of the lines we observed. The only exception is the J=$8_{-1,8}-7_{0,7}$ transition of \mtoh, where we are resolving out $\sim$90\% of the flux.

The correlator for the first track was configured to have two 500 MHz wide windows for continuum and four 32 MHz windows for spectral lines (half in each sideband), while the correlator for the second track was configured to have six 32 MHz windows (three in each sideband). The spectral line windows had 63 channels with a spacing of 488 kHz ($\sim$0.64 km s$^{-1}$). Uranus was used as the flux density calibrator and 0530+135 was used to calibrate the antenna based gains. The absolute amplitude calibration of 0530+135 from the flux density calibrator is accurate to within $\sim$20\%. The internal noise source and 0530+135 were used to correct the passbands of each window. The data were calibrated, continuum subtracted, and imaged using the MIRIAD software package \citep{sault95}.

\section{RESULTS}
Table~\ref{tab:mols} lists all the molecules that we searched for along with their parameters. The first column gives the molecular formula, the second column gives the transition quantum numbers, the third column gives the transition frequency in MHz, the fourth column gives the upper state energy ($E_u$) in K, the fifth column gives the line strength multiplied by the square of the relevant dipole moment ($S\mu^2$) in Debye$^2$, the sixth column gives the synthesized beam size from the observations in arcseconds, the seventh column gives the 1 $\sigma$ rms noise level from the associated spectral window in m\jbm, the eighth column given the percentage of missing flux when compared to the single dish observations of \citet{sutton87}, and the ninth column gives the literature reference for the molecular parameters. The table has been broken into 2 parts. The upper section lists the transitions detected in narrowband widows while the lower section lists transitions detected in wideband windows.

\subsection{Continuum}
\begin{figure}
\includegraphics[angle=270,scale=0.9]{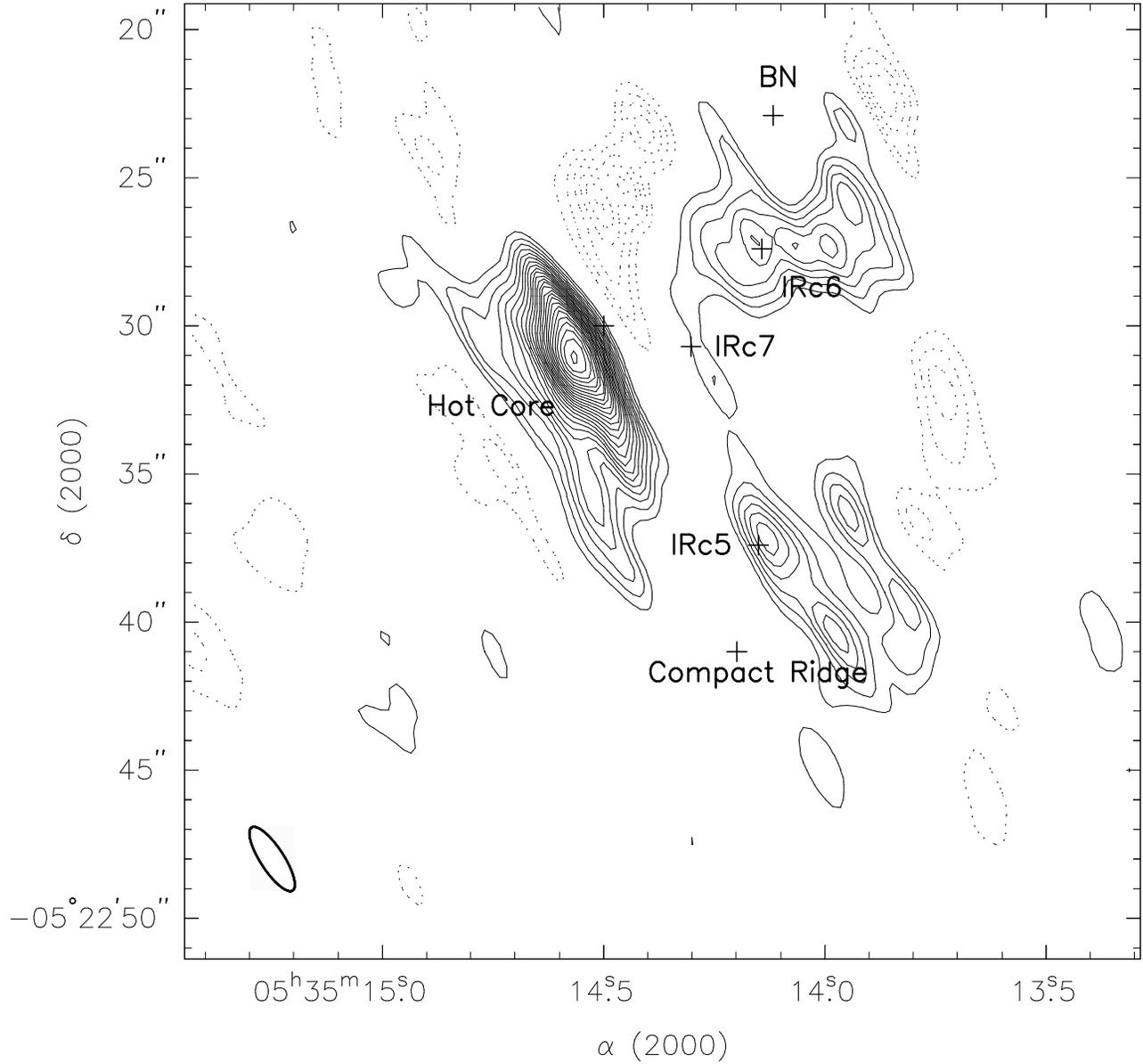}
\caption{Continuum of Orion-KL at 231 GHz. Known infrared and radio sources are denoted with ``+'' and labeled. Contour levels are $\pm3\sigma$, $\pm5\sigma$, etc., where the 1 $\sigma$ rms noise level is 11.3 m\jbm. The synthesized beam is in the lower left corner.\label{fig:contin}}
\end{figure}

We used two wideband ($\sim$500 MHz) windows to map the continuum of Orion-KL. Several molecular lines were detected in these windows (see \S\ref{sec:wideband}) and were removed before producing the final continuum image. Figure~\ref{fig:contin} shows the 231 GHz continuum of Orion-KL. Known bright infrared and radio sources\footnote{Some sources are not labeled as they are too close together, namely IRc2 and radio source I, near the hot core.} are denoted with a ``+'' and are labeled. All maps in this paper are contoured in $\pm3~\sigma$, $\pm5~\sigma$, etc. contours. The 1 $\sigma$ rms noise level of the continuum is 11.3 m\jbm\ (1 $\sigma$ rms noise levels for the other maps are listed in Table~\ref{tab:mols}). Most of the continuum comes from the hot core (865 m\jbm\ peak flux), however there is notable emission from IRc6 (232 m\jbm\ peak flux), and there is also emission associated with IRc5 (254 m\jbm\ peak flux) and IRc7 (101 m\jbm\ peak flux). No emission was detected from either the compact ridge or BN, above our 1 $\sigma$ rms noise level. There is, however, a notable hole in the continuum around the location of BN.

Previous 1mm array observations of this region by \citet{blake96} indicated that the hot core was composed of at least 2 clumps and high resolution observations of ammonia [NH$_3$] by \citet{migenes89} indicate numerous clumps along the long axis of the hot core. Additionally, \citet{beuther04} detected several continuum peaks near the hot core at 865 $\mu$m, which we were not able to resolve. The elongation of the continuum is also closely aligned with known SiO outflows seen by \citet{wright95}. Note that the emission from IRc6 is not a point source but rather a rough ``C'' shape around BN. There are also notable emission peaks both to the southwest and west of IRc5 which are not associated with known infrared sources.

\subsection{Methanol [\mtoh]}
\begin{figure}
\includegraphics[angle=270,scale=0.8]{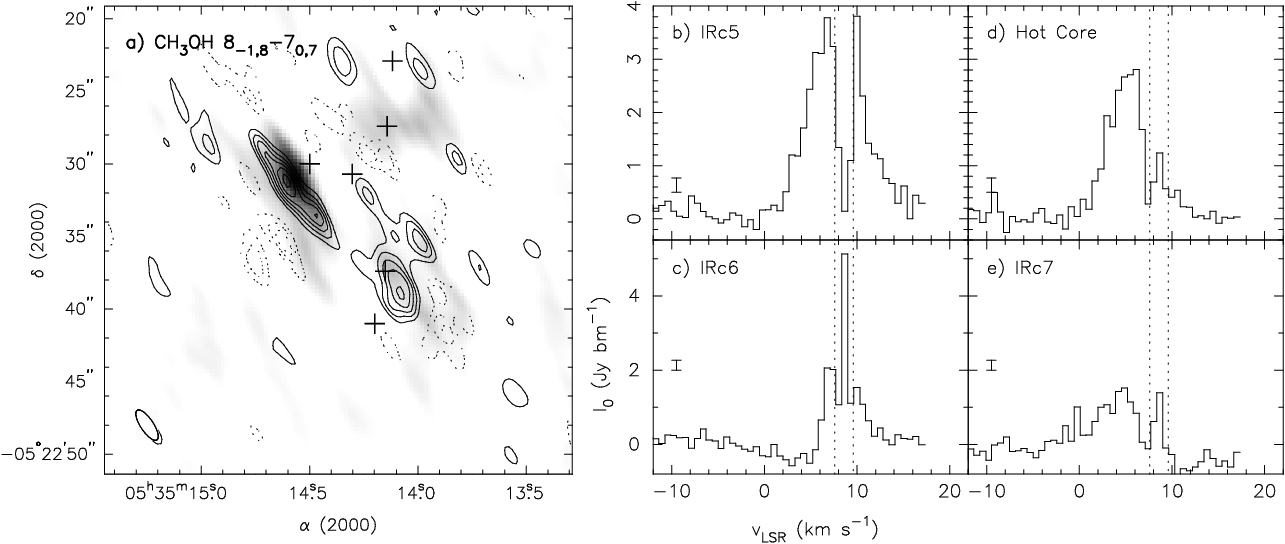}
\caption{Map and spectra of \mtoh. a) Map of the J=$8_{-1,8}-7_{0,7}$, \vlsr=5.5 \kms\ emission overlayed on gray-scale continuum. Contours are $\pm3\sigma$, $\pm5\sigma$, ... ($\sigma=0.268$ \jbm). The synthesized beam is in the lower left corner of the map and ``+'' marks denote the same sources as in the continuum map (see Figure~\ref{fig:contin}). b) Spectra toward IRc5. Dashed lines mark \vlsr=7.6 and 9.6 \kms. ``I'' bar denotes 1 $\sigma$ rms noise level. c) Spectra toward IRc6. d) Spectra toward hot core. e) Spectra toward IRc7.\label{fig:mtoh}}
\end{figure}

Two transitions of \mtoh\ were detected in our narrowband windows. Figure~\ref{fig:mtoh}a shows the channel map of the \vlsr=5.5 \kms\ emission of the lower energy (J=$8_{-1,8}-7_{0,7}$) transition of \mtoh. The \mtoh\ map indicates some wide spread emission but has distinct concentrations near the hot core, IRc5, IRc7, and IRc6 (some are not seen in the map as they are at different velocities), but possibly most notable is lack of emission from the compact ridge. There is also a notable emission peak extending to the southwest from the hot core (hereafter hot core-SW) that corresponds to weak continuum emission. Convolving our results with a 29\arcsec beam and comparing them to those from \citet{sutton87} indicates that we are resolving out a large percentage of flux $\sim$90\% from this transition.

Figures~\ref{fig:mtoh}b-e show the spectra toward IRc5, IRc6, the hot core, and IRc7 respectively. The highly asymmetric line profile of the IRc5 spectra indicates that \mtoh\ may be self absorbing near 9 \kms\ along this line of sight, while the IRc6 spectra peak at the same velocity as the IRc5 absorption. Both the hot core and IRc7 spectra show possible absorption features near 7.6 \kms.

Fits to the spectral lines are given in Table~\ref{tab:fits}. The first column lists the molecule and the second column lists the source. The third column gives the figure number for the spectra. The fourth, fifth, and sixth columns list the fit parameters: intensity in \jbm, $v_{\rm LSR}$ in \kms, and FWHM in \kms, respectively. The seventh and ninth columns list the calculated opacity at the line center for assumed lower and upper rotation temperatures respectively (see \S\ref{sec:trot} \& \S\ref{sec:NT}). The eighth and tenth columns list the total column density for the assumed lower and upper rotation temperatures in cm$^{-2}$ (see \S\ref{sec:NT}). The fits to the spectra are chi squared gaussian fits and only those lines for which we have a full spectral profile have been fit.

\begin{figure}
\includegraphics[angle=270,scale=0.9]{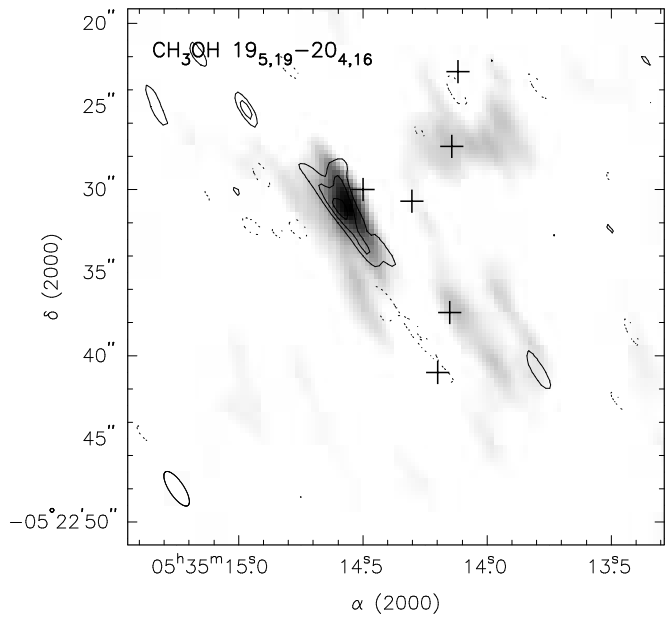}
\caption{Map of \mtoh\ J=$19_{5,15}-20_{4,16}$, \vlsr=5.4 \kms\ emission overlayed on gray-scale continuum. Contours are $\pm3\sigma$, $\pm5\sigma$, ... ($\sigma=0.188$ \jbm). The synthesized beam is in the lower left corner of the map and ``+'' marks denote the same sources as in the continuum map (see Figure~\ref{fig:contin}).\label{fig:mtoh2}}
\end{figure}

We did not set out to detect the second, narrowband, \mtoh\ transition but it was fortuitously at the edge of one of our spectral windows. Unfortunately we did not observe the entire line of this higher energy (J=$19_{5,15}-20_{4,16}$) \mtoh\ transition. We present a map of its emission at a \vlsr\ of 5.5 \kms\ in Figure~\ref{fig:mtoh2}. The emission is restricted to the hot core region and shows a slight extension to the southwest, similar to that of the other \mtoh\ transition.

\subsection{Dimethyl Ether [\dme]}
\begin{figure}
\includegraphics[angle=270,scale=0.9]{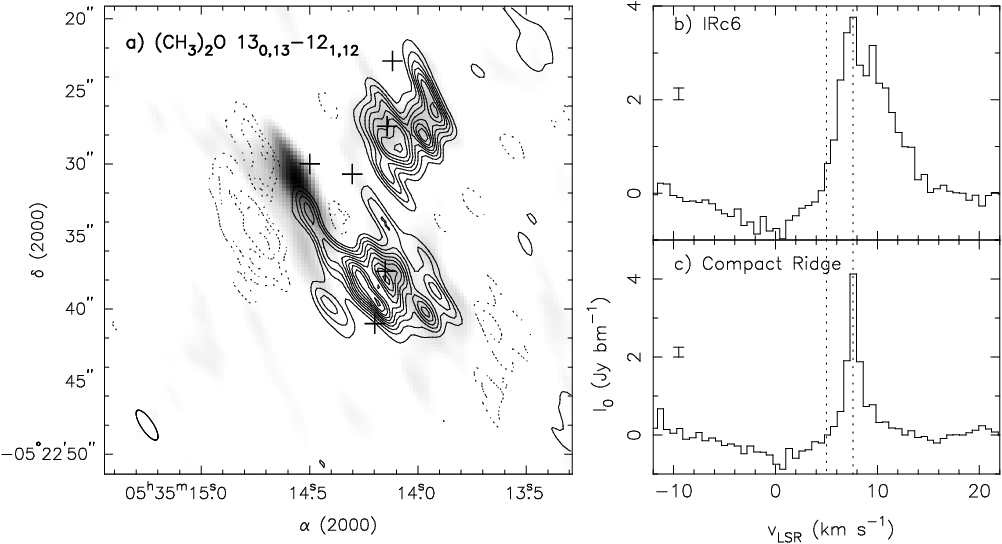}
\caption{Map and spectra of \dme. a) Map of the \vlsr=7.6 \kms\ emission overlayed on gray-scale continuum. Contours are $\pm3\sigma$, $\pm5\sigma$, ... ($\sigma=0.252$ \jbm). The synthesized beam is in the lower left corner of the map and ``+'' marks denote the same sources as in the continuum map (see Figure~\ref{fig:contin}). b) Spectra toward IRc6. Dashed lines mark 5.0 and 7.6 \kms\ (typical rest velocities for this region). ``I'' bar denotes 1 $\sigma$ rms noise level. c) Spectra toward compact ridge.\label{fig:dme}}
\end{figure}

\dme\ is a typical large O-bearing species and has been associated with the compact ridge \citep[e.g.][]{liu02}. Figure~\ref{fig:dme}a shows the \dme\ map at a \vlsr\ of 7.6 \kms. There clearly is some extended emission as indicated by the large negative contours. There is significant emission from IRc6 with a notable hole near Orion-BN, and in regions near IRc5. As with \mtoh\ there is no notable emission from the compact ridge itself. Also there is some emission from the hot core-SW. This emission is similar to that seen by \citet{beuther05} at 865 $\mu$m, however they do not see the ``C'' shape around BN, but rather see more point like emission from IRc6. This may be explained  by the factor of $\sim$2 difference in $E_u$ between the observed transitions. Figures~\ref{fig:dme}b and c show the \dme\ spectra from IRc6 and IRc5, respectively, and Table~\ref{tab:fits} shows the fits to the spectra. The IRc6 emission is best fit by two gaussians, one narrow the other wide. The wide absorption feature, near 0 \kms, seen in both spectra are due to the sidelobes of the \etcn\ (J=$27_{0,27}-26_{0,26}$) emission from the hot core.

\subsection{Methyl Formate [\mef]}
\begin{figure}
\includegraphics[angle=270,scale=0.9]{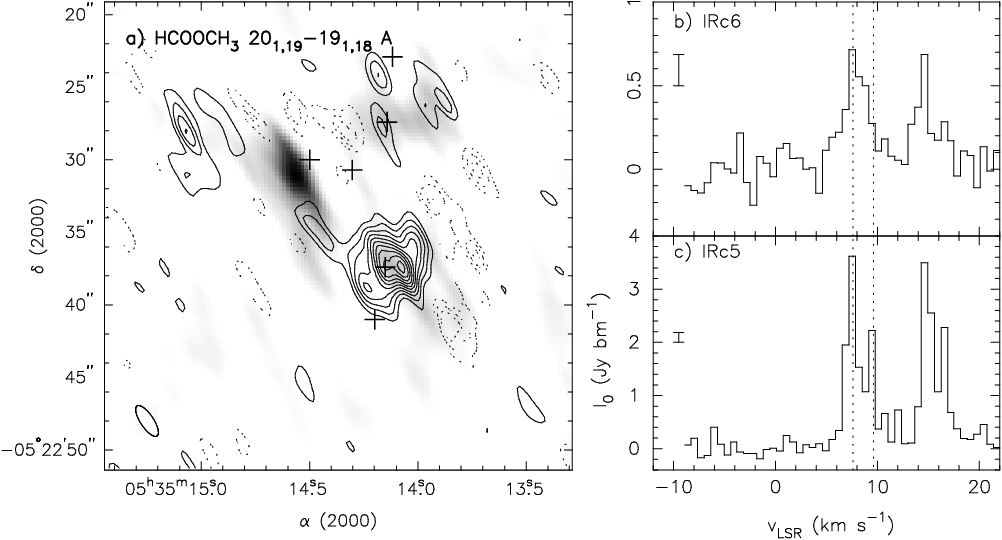}
\caption{Map and spectra of \mef. a) Map of the A torsional state, \vlsr=7.6 \kms\ emission overlayed on gray-scale continuum. Contours are $\pm3\sigma$, $\pm5\sigma$, ... ($\sigma=0.185$ \jbm). The synthesized beam is in the lower left corner of the map and ``+'' marks denote the same sources as in the continuum map (see Figure~\ref{fig:contin}). The emission component to the northeast of the hot core may be a sidelobe that was not completely removed. b) Spectra toward IRc6. Dashed lines mark \vlsr=7.6 and 9.2 \kms. ``I'' bar denotes 1 $\sigma$ rms noise level. The second set of spectral features near 15 \kms\ are the emission from the E torsional state of this transition.  c) Spectra toward IRc5.\label{fig:mef}}
\end{figure}

Similar to \dme, \mef\ has traditionally been associated with the compact ridge. Both the A and E torsional states of the \mef\ J=$20_{1,19}-19_{1,18}$ transition were detected. Figure~\ref{fig:mef}a shows the \vlsr=7.6 \kms\ map of the A torsional state. Unlike \dme, the strongest emission from \mef\ comes from the IRc5 region and not IRc6. Additionally there is some emission from the hot core-SW and no detectable emission from the compact ridge. As with \dme, there are some notable differences between our map and that observed by \citet{beuther05}, especially the low level of emission we observe toward IRc6. This difference may be explained by the large (nearly factor of 3) difference in $E_u$ of the observed transitions or by our single channel maps vs their velocity-integrated maps. The peak in the northeast of our map is most likely from sidelobes that were not completely removed. Figures~\ref{fig:mef}b and c show the \mef\ spectra toward IRc6 and IRc5, respectively. The velocity scale is in reference to the A torsional state transition. The IRc5 emission appears to have two peaks at $\sim$7.6 and 9.3 \kms.

\subsection{Ethyl Cyanide [\etcn]}
\begin{figure}
\includegraphics[angle=270,scale=0.9]{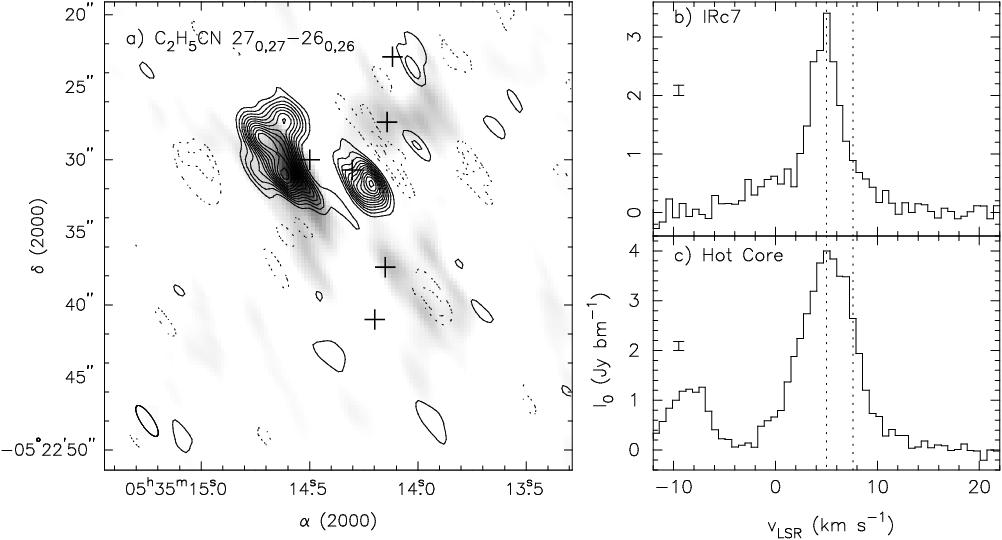}
\caption{Map and spectra of \etcn. a) Map of the J=$25_{3,22}-24_{3,21}$, \vlsr=5.0 \kms\ emission overlayed on gray-scale continuum. Contours are $\pm3\sigma$, $\pm5\sigma$, ... ($\sigma=0.179$ \jbm). The synthesized beam is in the lower left corner of the map and ``+'' marks denote the same sources as in the continuum map (see Figure~\ref{fig:contin}). b) Spectra toward IRc7. Dashed lines mark \vlsr=5.0 and 7.6 \kms\ (typical rest velocities for this region). ``I'' bar denotes 1 $\sigma$ rms noise level. c) Spectra toward the hot core.\label{fig:etcn}}
\end{figure}

\etcn\ is a typical large N-bearing species and has been associated with the hot core for many years \citep[e.g.][]{blake87}. \citet{blake96} and \citet{beuther05} showed an extension in \etcn\ emission toward IRc7. The map of \etcn\ J=$25_{3,22}-24_{3,21}$ emission, shown in Figure~\ref{fig:etcn}a, clearly resolves this second component for the first time. It also shows an extended ridge of emission running northeast-southwest along the hot core and in the direction of known outflows. Additionally, there is a peak of emission to the north of the hot core, a region which has no detected continuum emission. The emission from IRc7 may be coming from two (or more) partially resolved sources. Figures~\ref{fig:etcn}b and c show the \etcn\ spectra toward IRc7 and the hot core, respectively. The \etcn\ spectra from IRc7 were best fit by two gaussians, one narrow and one wide, while the hot core spectra were best fit by a single gaussian. The second line (\vlsr$\sim$-9.5 \kms) in the hot core spectra may be a U-line (unidentified) or it may be a velocity component of \etcn. This line does not have a large spatial distribution and is confined to the southwest corner of the hot core. The other transition of \etcn\ we observed (J=$27_{0,27}-26_{0,26}$) maps nearly identically to this transition.

\begin{figure}
\includegraphics[angle=270,scale=0.9]{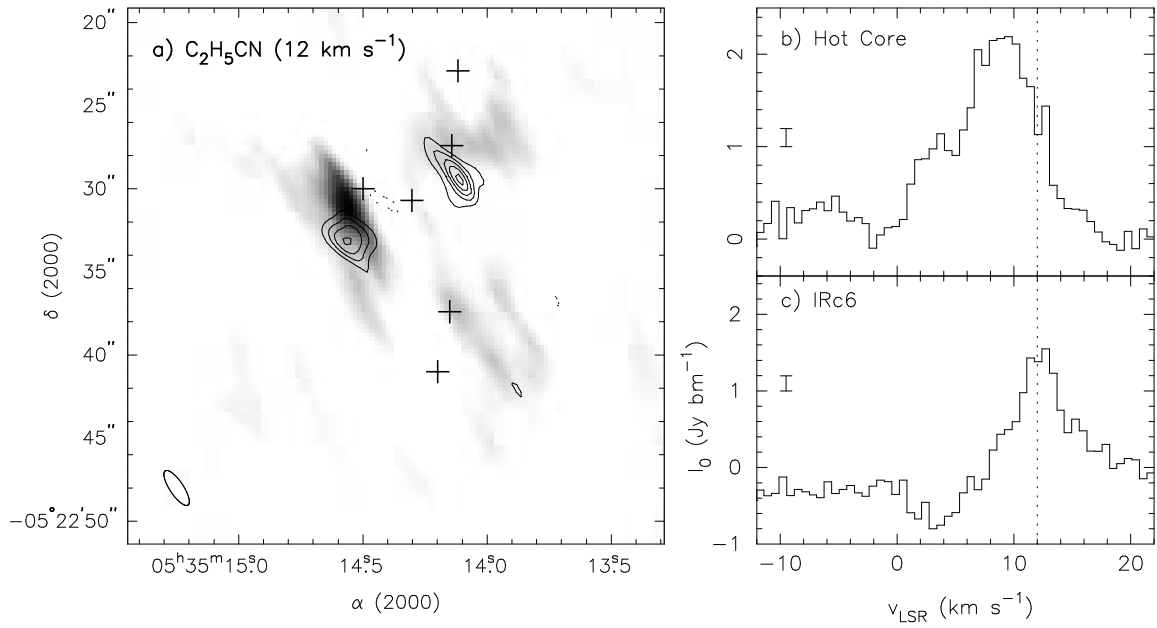}
\caption{Map of \etcn\ at a \vlsr\ of 12 \kms. a) Contours are the same as in Figure~\ref{fig:etcn}. Note the emission south of IRc6. b) Spectra toward the emission peak south of the hot core. c) Spectra toward the emission peak near IRc6.\label{fig:etcn12}}
\end{figure}

One notable feature of the \etcn\ emission is a strong peak at a \vlsr\ of 12 \kms\ just south of IRc6. Figure~\ref{fig:etcn12}a shows the channel map of this emission and Figures~\ref{fig:etcn12}b and c show the spectra from near the hot core and IRc6 respectively. There is no notable continuum associated with this emission and nothing else like it appears in any of the other transitions we observed.

\begin{figure}
\includegraphics[angle=270,scale=0.9]{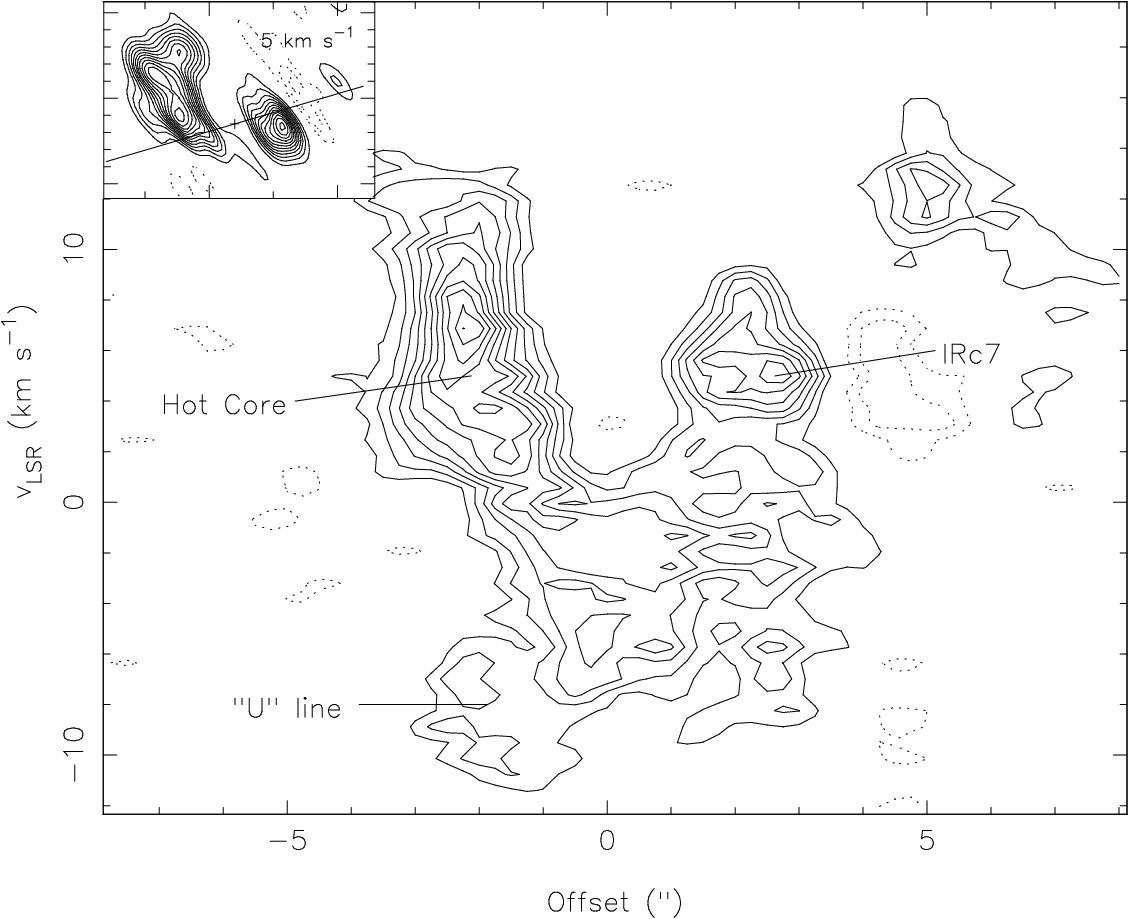}
\caption{Position-velocity diagram of \etcn. The inset is a zoom in on the hot core/IRc7 emission, at a \vlsr\ of 5 \kms, shown in Figure~\ref{fig:etcn}. The line denotes the p-v cut and the ``+'' denotes its center ($\alpha{\rm(J2000)}=05^h35^m14^s.4$, $\delta{\rm(J2000)}=-5{\degr}22{\arcmin}31{\arcsec}.5$). The abscissa of the diagram is offset arcseconds from the p-v center and the ordinate is $v_{\rm LSR}$ in \kms. Labels indicate the position of the hot core and IRc7 along with the potential U-line. The contours are the same as those in Figure~\ref{fig:etcn}\label{fig:pv}}
\end{figure}

While there is no notable velocity gradient along the ridge of the hot core there are some velocity gradients across the hot core near its southwestern tip. Figure~\ref{fig:pv} shows a position-velocity (p-v) diagram of \etcn. The inset is a map of \etcn\ emission at a \vlsr\ of 5 \kms\ (a zoom in of Figure~\ref{fig:etcn}a). The line denotes the p-v cut and the ``+'' denotes the center position of the cut ($\alpha{\rm(J2000)}=05^h35^m14^s.4$, $\delta{\rm(J2000)}=-5{\degr}22{\arcmin}31{\arcsec}.5$). The abscissa of the p-v diagram is offset arcseconds from the center point and the ordinate is $v_{\rm LSR}$ in \kms. The positions of the hot core and IRc7 are noted in the diagram as is the position of the U-line. The contours are the same as in Figure~\ref{fig:etcn}. The hot core contains not only \etcn\ gas at 5.0 \kms\ but some higher velocity gas, near a \vlsr\ of 7.5 \kms\ (the typical compact ridge velocity). The diagram also shows that the hot core and IRc7 are connected by a ridge of lower velocity gas. This ridge also connects with the U-line indicating that it might be \etcn\ at a lower velocity. The U-line may also be the $^1\nu_6$ J=$25-24~1e$ transition of HC$_3$N at 227.7916 GHz \citep{cdms}, a rather high energy line ($E_u$=859 K), which could explain its compactness.

\subsection{Acetone [\acetone]}
\begin{figure}
\includegraphics[angle=270,scale=0.9]{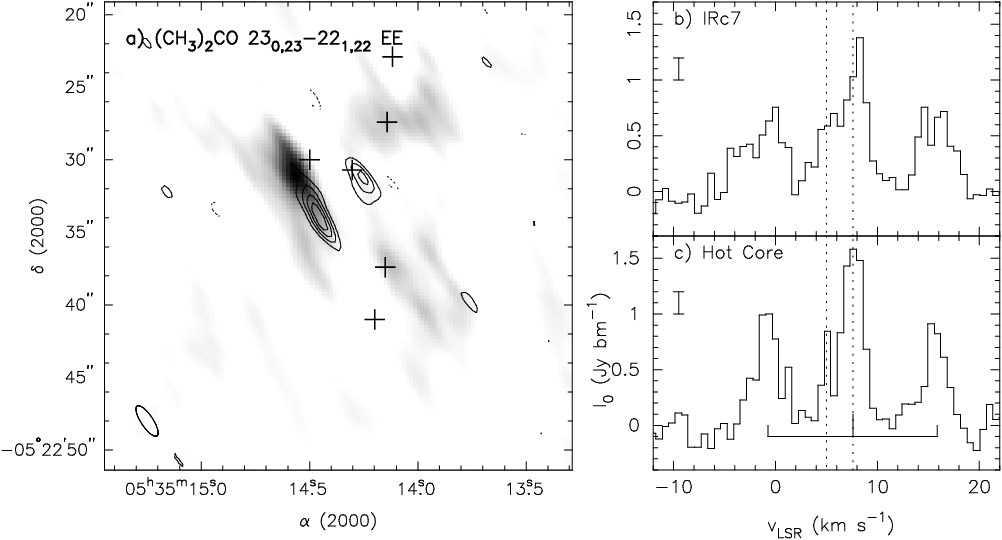}
\caption{Map and spectra of \acetone. a) Map of the J=$23_{0,23}-22_{1,22}$ EE, \vlsr=5.0 \kms\ emission overlayed on gray-scale continuum. Contours are $\pm3\sigma$, $\pm5\sigma$, ... ($\sigma=0.196$ \jbm). The synthesized beam is in the lower left corner of the map and ``+'' marks denote the same sources as in the continuum map (see Figure~\ref{fig:contin}). b) Spectra toward IRc7. Dashed lines mark \vlsr=5.0 and 7.6 \kms\ (typical rest velocities for this region). ``I'' bar denotes 1 $\sigma$ rms noise level c) Spectra toward hot core.\label{fig:ace}}
\end{figure}

\acetone, while a large O-bearing species, does not map like any of the other molecules we observed. Figure~\ref{fig:ace}a shows a map of the \acetone\ emission. Two distinct emission features are seen. One at hot core-SW, the other near IRc7. Figures~\ref{fig:ace}b and c show the spectra toward IRc7 and the hot core, respectively, and Table~\ref{tab:fits} lists the parameters of the spectral fits. All three spectral features (AE/EA, EE, \& AA) are seen at both positions. Both spectra are best fit by two velocity components, $\sim$5 \kms\ and $\sim$8 \kms, typical $v_{\rm LSR}$ values for the hot core and compact ridge. This is the largest known molecule that shows both velocity components, which are co-spatial at our resolution. Unlike \etcn\ there are no velocity gradients in any of the \acetone\ emission.

\subsection{Formic Acid [\fa], Acetaldehyde [\acal], and Ethanol [\etoh]}
The \fa, \acal, and \etoh\ transitions listed in Table~\ref{tab:mols} were not detected above their respective 3 $\sigma$ detection limits. This may be due to these molecules having a generally extended emission and thus being resolved out by the array, or the transitions we observed were not favorable, or both.

\subsection{Lines Detected in Wideband Windows}\label{sec:wideband}
\begin{figure}
\includegraphics[angle=270,scale=0.9]{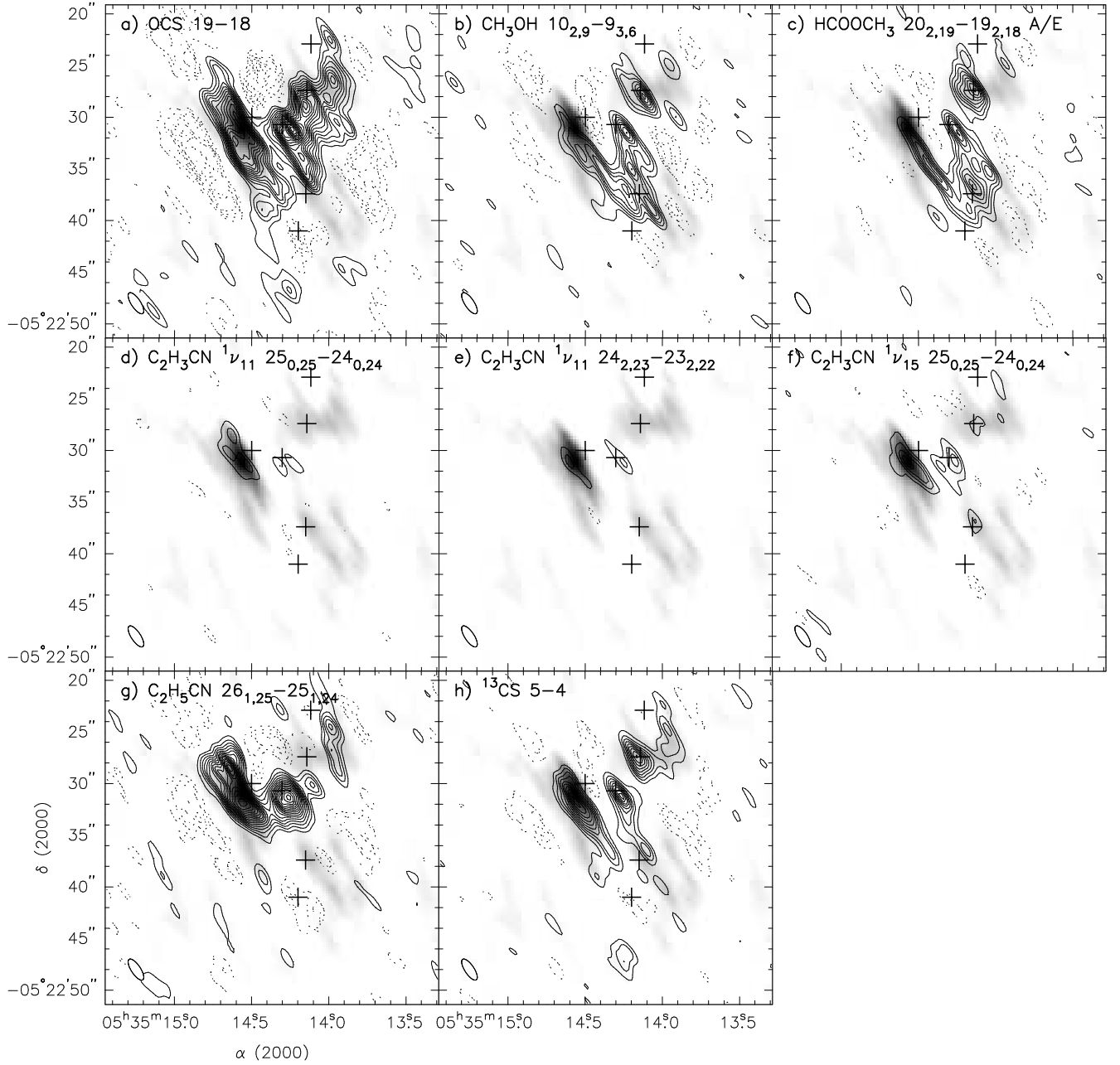}
\caption{Maps from the transitions detected in the wideband windows overlayed on gray-scale continuum. Contours are $\pm3\sigma$, $\pm5\sigma$, ... (1 $\sigma=29$ m\jbm). The synthesized beam is in the lower left corner of the map and ``+'' marks denote the same sources as in the continuum map (see Figure~\ref{fig:contin}).\label{fig:multi}}
\end{figure}

Several molecular transitions were strong enough to detect in our wideband windows ($\sim$500 MHz bandwidth, $\sim$ 41.3 \kms\ resolution). Ten of the twelve transitions from six molecules detected in the wideband channels are listed in the lower part of Table~\ref{tab:mols}\footnote{The other two transitions are from \mef\ and were also detected in the narrow spectral windows}. Figure~\ref{fig:multi} shows the channel maps of the transitions. The 1 $\sigma$ rms noise level for all maps in this figure is 29 m\jbm. Note that all maps in this figure are one channel wide, and cover all the velocity components of the transition.

Figure~\ref{fig:multi}a shows the OCS J=19-18 emission. There are notable peaks and valleys all throughout the region indicating widespread features that are resolved out by the array, in addition to the smaller scale structures that are seen. For such a low temperature transition from a simple molecule this is not a surprising result. Figure~\ref{fig:multi}b shows the \mtoh\ J=$10_{2,9}-9_{3,6}$ emission. The emission is very similar to that of the \mtoh\ J=$8_{-1,8}-7_{0,7}$ transition.

Figure~\ref{fig:multi}c shows the \mef\ J=$20_{2,19}-19_{2,18}$ A and E emission, which maps very similarly to the J=$20_{1,19}-19_{1,18}$ A and E emission, with the exception that these \mef\ transitions have a strong peak near IRc6, which may indicate contamination from another species in the wideband channel or may be due to velocity-averaging (similar to that of \citet{beuther05}). Figure~\ref{fig:multi}d-f show the $^1\nu_{11}$ J=$25_{0,25}-24_{0,24}$ and J=$24_{2,23}-23_{2,22}$ and $^1\nu_{15}$ J=$25_{0,25}-24_{0,24}$ transitions of vinyl cyanide [\vycn], respectively. These are the only transitions of \vycn\ detected and all are from vibrationally excited states. They map very similarly to the \etcn\ transitions, having strong emission from the hot core region and IRc7. One notable difference is that the \vycn\ emission shows two emission peaks in IRc7 while \etcn\ only hints at two spatially blended components. These two components have peaks at $\alpha$(J2000) = $05^h35^m14^s.333$, $\delta$(J2000) = $-05{\degr}22{\arcmin}31{\arcsec}.5$ and $\alpha$(J2000) = $05^h35^m14^s.239$, $\delta$(J2000) = $-05{\degr}22{\arcmin}31{\arcsec}.1$.

Figure~\ref{fig:multi}g shows the J=$26_{1,25}-25_{1,24}$ transition of \etcn. While this transition maps similarly to the other detected \etcn\ transitions there is one significant difference. The emission near IRc6 seen here is not detected in any other \etcn\ transitions. This may be due to contamination from weaker lines of O-bearing species (several \acal\ transitions fell within the same channel). Lastly, Figure~\ref{fig:multi}h shows the J=5-4 transitions of $^{13}$CS. The map shows strong emission from the hot core, IRc7, and IRc6 and weaker emission from IRc5 and the hot core-SW.

\section{DISCUSSION}
\subsection{Sources}
\subsubsection{Hot Core and IRc7}
As can been seen from the maps of \etcn\ and \vycn, when observed under high spatial resolution, molecules that traditionally have been associated with the hot core have multiple strong components: one or two sources associated with IRc7 and a string of components associated with the hot core.

The spectral line fits to the hot core features (listed in Table~\ref{tab:fits}) show a common $\sim$5.2 \kms\ velocity component (the canonical rest velocity for this source) for all species detected there. However, the two O-bearing species detected toward the hot core/hot core-SW, \mtoh\ and \acetone, have a second, narrower, component near 7.5 \kms (a typical rest velocity for compact ridge species). How this second component is associated with the hot core is unclear, although since the \mtoh\ line is in absorption it is likely that this component is from gas along the line of sight.

The spectral line fits to the IRc7 features have fewer commonalities. The \etcn\ spectrum is best fit by a pair of gaussians at $\sim$4 and $\sim$5 \kms, both with distinctly different line widths. The O-bearing species in IRc7 also have 2 components, one near 5.1 \kms\ the other near 7.5 \kms. The latter velocity component is in absorption in \mtoh, indicating that it is cooler and may be between us and IRc7. The \vycn\ maps give a strong indication of multiple cores which may have differing velocities. However they are unresolved in all other lines, so future observations will be needed to study these further.

\subsubsection{Compact Ridge, IRc6, and IRc5}
The spectral line fits to all the IRc6 features have a common component near 7.4 \kms. Additionally \dme\ and \mtoh\ have secondary velocity components at $\sim$9-10 \kms. As none of the \mtoh\ components are in absorption it is likely that all velocity components are warmer. Also it is evident in the maps that there are multiple unresolved components in IRc6, each of which may have a slightly different \vlsr. Comparing our maps with those presented by \citet{beuther05} indicates that the structure that is observed in IRc6 may be more dependent on the $E_u$ of the transition observed, than it is for any of the other sources. This indicates that IRc6 may be at a notably different temperature, or have distinctly different physical conditions, than the other regions.

As with IRc6, IRc5 has two distinct velocity components, one near 7.6 \kms, the other near 8.5 \kms, for all transitions observed toward this source. The \mtoh\ transition is in absorption, indicating that it may be between IRc5 and us. As with our other sources, IRc5 appears to have at least 2 partially resolved components, which may be the sources of the different velocity components. As no transitions of larger molecules were detected toward the compact ridge it is likely that many molecules typically associated with the compact ridge, via their \vlsr, may actually be in IRc5 instead. One exception that we know of is \fa. Observations comparing the spatial distribution of \dme\ and \fa\ by \citet{liu02} show that \fa\ may be associated with the compact ridge as its emission is to the southeast of that of \dme\ and appears to be on top of the compact ridge.

\subsection{Rotation Temperature Estimation}\label{sec:trot}
As we did not detect enough transitions of any one molecule to construct a reasonable rotation-temperature diagram we will assume a range of temperatures for the different molecular species. From the 3mm spectral line survey of Friedel et al., in preparation, we will use temperatures of 100-150 K for O-bearing (compact ridge) species and 200-500 K for N-bearing (hot core) species. Note that for \acetone\ we assume a temperature range of the other O-bearing species as its original detection gave  rotation temperature of $\sim$176 K, notably below the range from the hot core.

\subsection{Opacity and Column Densities}\label{sec:NT}
High optical depth ($\tau>1$) can greatly reduce the observed intensity of a spectral line. From \citet{rw00} the opacity at the line center can be calculated from
\begin{equation}
\tau=-ln\left[1-\frac{T_{MB}}{J(T_{r})-J(2.73~{\rm K})-T_C}\right]\label{eqn:tau},
\end{equation}
where $T_{MB}=T_R^*/\eta_M$ is the main beam brightness temperature of the line and $\eta^*_M$ is the corrected main beam efficiency. $J(T)$ is defined as \citep{rw00}
\begin{equation}
J(T)=\frac{h\nu}{k}\frac{1}{e^{h\nu/kT}-1}~{\rm K},
\end{equation}
$J(2.73)$ is the cosmic microwave background, and $T_C$ is the continuum brightness temperature. For observations not done in a temperature scale one can convert from Jy beam$^{-1}$ to K using \citep{rw00}
\begin{equation}
T_{MB}=\frac{1.22I_0}{B\theta_a\theta_b\nu^2}\times 10^6~{\rm K},
\label{eqn:i2t}
\end{equation}
where $I_0$ the is peak intensity of the line in Jy beam$^{-1}$, $B$ is the beam filling factor (assumes to be 0.5, filling the beam, for all calculations here), $\theta_a$ and $\theta_b$ are the FWHM Gaussian synthesized beam axes in arcseconds, and $\nu$ is the transition frequency in GHz. For emission lines the optical depth correction factor, $C_\tau$ can then be calculated \citep{goldsmith99}:
\begin{equation}
C_\tau=\frac{\tau}{1-e^{-\tau}}.
\label{eqn:ct}
\end{equation}
This factor is then applied to the column density equation~(\ref{eqn:nta}). The opacities for the detected transitions are given in the seventh (low temperature limit, 150 K for hot core/IRc7 and 100 K for all others) and ninth columns (high temperature limit, 500 K for hot core/IRc7 and 150 K for all others) of Table~\ref{tab:fits}.

 With the assumption of LTE, to calculate the total beam averaged column density ($\langle N_T\rangle$) from a single
 transition for an array, we have \citep[e.g.,][]{friedel05,remijan05}
\begin{equation}
\frac{N_u}{g_u}=\frac{2.04WC_\tau}{B\theta_a\theta_bS\mu^2\nu^3}\frac{T_r}{T_r-T_b}\times10^{20}~{\rm cm}^{-2}
\label{eqn:nta}
\end{equation}
and
\begin{equation}
\langle N_T\rangle=\frac{N_u}{g_u}Q_{rv}e^{E_u/T_{r}}.
\label{eqn:nunt}
\end{equation}
Here $W$ is the integrated line intensity in Jy beam$^{-1}$ km s$^{-1}$, $Q_{rv}$ is the rotational-vibrational partition function \citep{friedel05,ww05}, $E_u$ is the upper state energy of the transition in K, $T_r$ is the rotational temperature in K, $S\mu^2$ is the product of the line strength and the square of the relevant dipole moment in Debye$^2$, $N_u$/$g_u$ is the upper state column density divided by its degeneracy (2$J$+1), and $T_b$ is the total background continuum temperature. The column densities for each transition are given in the eighth (low temperature limit, 150 K for hot core/IRc7 and 100 K for all others) and tenth (high temperature limit, 500 K for hot core/IRc7 and 150 K for all others) columns of Table~\ref{tab:fits}.

From Table~\ref{tab:fits} it can be see that most of our lines are fairly optically thin at our low temperature limit. However, several transitions, notably those of O-bearing species, are optically thick. At our high temperature limit it can be seen that more transitions are optically thin when compared to the low temperature limit. In general the column densities for all transitions are notably higher than those reported previously in the literature. The reasons for this are 1) we are resolving the sources and thus do not suffer from beam dilution and 2) we are accounting for the vibrational part of the partition function which can increase $\langle N_T\rangle$ by factors of 5 or more.

\subsection{Implications for Astrochemistry}
One of the big questions about the Orion-KL region is what is causing the O-N differentiation. One could argue that the O-N differentiation seen in Orion-KL is due to the differing dipole moments (and hence differing critical densities) of O and N bearing molecules (in general N bearing molecules have notably larger dipole moments). In order to investigate this possibility we calculate the critical density for the \etcn\ and \dme\ transitions we detected. The critical densities, $n_{cr}$, can be calculated as follows \citep{tielens05}:
\begin{equation}
n_{cr}=\frac{\sum_{l<u}A_{ul}}{\sum_{l\ne u}\gamma_{ul}}
\label{eqn:ncrit}
\end{equation}
where $A_{ul}$ is the Einstein A of the transitions and $\gamma_{ul}$ is the collisional de-excitation rate coefficient. Since the quantum levels of the \etcn\ and \dme\ transitions are dominated by single radiative routes in and out, we can consider this to be a two level system and the summation collapses to single expressions. $A_{ul}$ is defined as:
\begin{equation}
A_{ul} = \frac{64\pi^4}{3hc^3}\nu^3\frac{S\mu^2}{g_u},
\label{eqn:einstein}
\end{equation}
and we find for the observed transitions of \etcn\ (J=$27_{0,27}-26_{0,26}$) and \dme\ (J=$13_{0,13}-12_{1,12}$), $A_{ul}$=9.8$\times10^{-4}$ and 9.1$\times10^{-5}$ s$^{-1}$ respectively. The de-excitation coefficient is defined as \citep{tielens05}:
\begin{equation}
\gamma_{ul}=\frac{4}{\sqrt{\pi}}\left(\frac{\mu}{2kT}\right)^{3/2}\int_0^\infty\sigma_{ul}(v)v^3e^{-\mu v^2/2kT}dv,
\end{equation}
where $\mu$ is the reduced mass of the system and $\sigma_{ul}(v)$ is the collisional de-excitation cross section at the relative velocity $v$. If we assume the geometrical cross section\footnote{This assumption is valid since the primary collision partner, H$_2$, has no permanent dipole moment and all other interactions are very short range, thus the collisional cross section will be dominated by the geometrical cross section.} of each molecule ($\sigma_{ul}(v)\approx1.3\times10^{-15}$ cm$^2$ for \etcn\ and $\sigma_{ul}(v)\approx7.1\times10^{-16}$ cm$^2$ for \dme), the primary collision partner to be H$_2$, and temperatures of 200 K and 100 K\footnote{Changing the assumed temperature will not affect the outcome much as $n_{cr}$ will go as $T^{-1/2}$ (higher temperature gives a lower $n_{cr}$).} for \etcn\ and \dme\ respectively, then we find $\gamma_{ul}=1.9\times10^{-10}$ cm$^3$s$^{-1}$ for \etcn\ and $\gamma_{ul}=7.4\times10^{-11}$ cm$^3$s$^{-1}$ for \dme. Using equation~(\ref{eqn:ncrit}) gives critical densities of 5.3$\times10^6$ cm$^{-3}$ for \etcn\ and 1.2$\times10^{6}$ cm$^{-3}$ for \dme. This slight difference in critical densities is unlikely to be the cause of the differentiation.

\begin{figure}
\includegraphics[angle=270,scale=0.9]{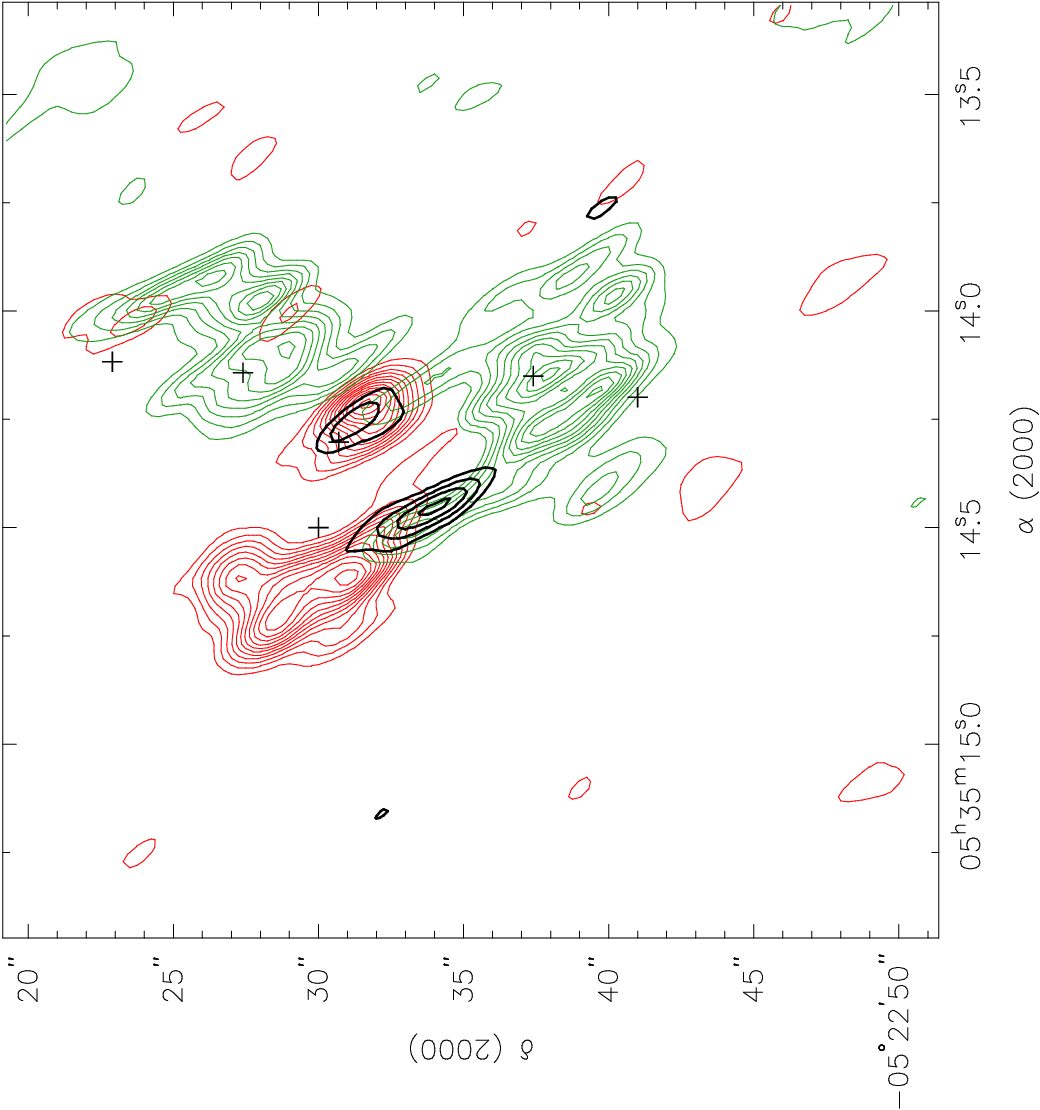}
\caption{Overlayed maps of \etcn\ in grey (red in electronic edition) contours, \dme\ in dashed (green in electronic edition) contours, and \acetone\ in black contours. Note that negative contours have been removed to reduce confusion. The \acetone\ emission only appears where both \etcn\ and \dme\ emission overlap.\label{fig:compare}}
\end{figure}

Others have argued that the differentiation could be due to differences in the age of the regions, the O-bearing species are liberated off grain surfaces by shocks or form in gas phase reactions in shocked regions, while the N-bearing species take longer to form in gas-grain reactions \citep[e.g.][]{blake87}. This argument fits well with what is suspected about the region: outflows from the older (N-bearing) hot core are impacting the ambient gas and dust forming shocked regions where larger O-bearing species are most populous. However, our observations of \acetone\ show that this O-N differentiation is not as simple as it may seem. Figure~\ref{fig:compare} shows \etcn\ in grey (red in electronic edition) contours, \dme\ as dashed (green in electronic edition) contours, and \acetone\ as black contours, overlayed (note that negative contours have been removed to reduce confusion). It shows that the \acetone\ emission comes only from regions where both \dme\ and \etcn\ overlap. If we assume that the age difference argument is correct then it would seem that it takes significant processing (either on grain surfaces or in the gas phase) to form \acetone, but that it is also short lived as it seems to not be present in older regions where N-bearing molecules dominate. Recent models by Garrod, Widicus Weaver, \& Herbst, in preparation, indicate that this may be true: \acetone\ forms where the warm-up time scale is longer, indicating grain surface formation from reactions with secondary products.

\section{SUMMARY}
We have conducted high resolution $\lambda$=1mm observations of the Orion-KL region with the CARMA array. Our observations studied the continuum of the region as well as eleven molecular species. Of these eleven species, eight were detected and we have presented their maps and spectra. These maps show that this region more complex than previously thought. The large N-bearing species (\etcn\ and \vycn) originate from multiple cores near the Orion hot core and IRc7, while large O-bearing species (\dme\ and \mef) originate near IRc5 and IRc6, and not the compact ridge as had been assumed before. Additionally we see that \acetone, unlike other large O-bearing species, does not originate from IRc5 or IRc6, but rather from the hot core-SW and IRc7, the only regions where both large N-bearing and O-bearing species are present.

\acknowledgements
We would like to thank S. Widicus Weaver and L. Looney for many helpful comments and discussions. We also thank an anonymous referee for a very favorable review of this work. This work was partially funded by NSF grant AST-0540459 and the University of Illinois. Support for CARMA construction was derived from the states of Illinois, California, and Maryland, the Gordon and Betty Moore Foundation, the Kenneth T. and Eileen L. Norris Foundation, the Associates of the California Institute of Technology, and the National Science Foundation. Ongoing CARMA development and operations are supported by the National Science Foundation under a cooperative agreement, and by the CARMA partner universities.

\clearpage
\begin{deluxetable}{lllrrrccc}
\tablecolumns{9}
\tabletypesize{\scriptsize}
\tablewidth{0pt}
\tablecaption{Molecular Parameters of Observed Lines}
\tablehead{\colhead{} & \colhead{Quantum} & \colhead{Frequency} & \colhead{$E_u$} & \colhead{S$\mu^2$} & \colhead{$\theta_a\times\theta_b$} & \colhead{RMS Noise} & \colhead{Missing} & \colhead{}\\
\colhead{Molecule} & \colhead{Numbers} & \colhead{(MHz)} & \colhead{(K)} &\colhead{($D^2$)}& \colhead{($\arcsec\times\arcsec$)} & \colhead{(m\jbm)} & \colhead{Flux (\%)\tablenotemark{a}} & \colhead{Ref.}}
\startdata
\cutinhead{Primary targets in narrowband windows}
\mef & $20_{1,19}-19_{1,18}$ A & 226,778.764 (10) & 120 & 51.4 & $2.49\times0.90$ & 185 & $\sim15$ & 1\\
& $20_{1,19}-19_{1,18}$ E & 226,773.152 (9) & 120 & 51.4 & $2.49\times0.90$ & 185 & $\sim15$ & 1\\
\dme & $13_{0,13}-12_{1,12}$ AA\tablenotemark{b} & 231,987.779 (3) & 81 & 16.9 & $2.42\times0.82$ & 252 & $\sim10$ & 2 \\
& $13_{0,13}-12_{1,12}$ EE\tablenotemark{b} & 231,987.856 (3) & 81 & 16.9 & $2.42\times0.82$ & 252 & $\sim10$ & 2 \\
& $13_{0,13}-12_{1,12}$ AE/EA\tablenotemark{b} & 231,987.933 (3) & 81 & 16.9 & $2.42\times0.82$ & 252 & $\sim10$ & 2 \\
\etcn & $27_{0,27}-26_{0,26}$ & 231,990.409 (3) & 157 & 399.4 & $2.50\times0.83$ & 252 & $<5$\tablenotemark{c} & 3 \\
& $25_{3,22}-24_{3,21}$ & 227,780.971 (2) & 150 & 365.4 & $2.42\times0.83$ & 179 & $<5$\tablenotemark{c} & 3 \\
\acetone & $23_{0,23}-22_{1,22}$ AE\tablenotemark{d} & 230,170.265 (20) & 135 & 22.4  & $2.47\times0.82$ & 196 & \tablenotemark{e} & 4 \\
 & $23_{0,23}-22_{1,22}$ EA\tablenotemark{d} & 230,170.313 (17) & 135 & 22.4  & $2.47\times0.82$ & 196 & \tablenotemark{e} & 4 \\
 & $23_{0,23}-22_{1,22}$ EE & 230,176.727 (14) & 135 & 22.4  & $2.47\times0.82$ & 196 & \tablenotemark{e} & 4 \\
 & $23_{0,23}-22_{1,22}$ AA & 230,183.083 (21) & 135 & 22.4  & $2.47\times0.82$ & 196 & \tablenotemark{e} & 4 \\
\mtoh & $8_{-1,8}-7_{0,7}$ & 229,758.811 (15) & 89 & 5.0 & $2.45\times0.88$ & 268 & $\sim90$ & 5\\
 & $19_{5,15}-20_{4,16}$ & 229,864.221 (46) & 578 & 5.7 & $2.43\times0.88$ & 188 & \tablenotemark{f} & 5\\
\fa & $10_{1,9}-9_{1,8}$ & 231,505.705 (23) & 64 & 19.3 & $2.38\times0.89$ & 279 & \tablenotemark{g} & 3\\
\acal & $11_{1,11}-10_{0,10}$ & 229,775.029 (11) & 61 & 8.4 &$2.45\times0.88$ & 284 & \tablenotemark{g} & 6\\
\etoh & $10_{2,9}-9_{1,8}$ & 226,661.701 (100) & 51 & 17.0 & $2.47\times0.90$ & 248 & \tablenotemark{g} & 3\\
\cutinhead{Transitions detected in wideband windows}
OCS & 19-18 & 231,060.983 (10) & 110 & 9.7 & $2.44\times0.83$ & 29 & \tablenotemark{h} & 3\\
\mtoh & $10_{2,9}-9_{3,6}$ & 231,281.150 (15) & 165 & 2.67 & $2.44\times0.83$ & 29 & \tablenotemark{h} & 5\\
\mef & $20_{2,19}-19_{2,18}$ A & 226,713.028 (9) & 120 & 51.4 & $2.48\times0.84$ & 29 & \tablenotemark{h} & 1\\
 & $20_{2,19}-19_{2,18}$ E & 226,718.748 (10) & 120 & 51.4 & $2.48\times0.84$ & 29 & \tablenotemark{h} & 1\\
\vycn\ $^1\nu_{11}$ & $24_{2,23}-23_{2,22}$ & 226,835.668 (16) & 488 & 346.7 & $2.48\times0.84$ & 29 & \tablenotemark{h} & 3\\
 & $24_{1,23}-23_{1,22}$ & 231,101.164 (14) & 486 & 348.3 & $2.44\times0.83$ & 29 & \tablenotemark{h} & 3\\
 & $25_{0,25}-24_{0,24}$ & 231,145.707 (14) & 490 & 363.1 & $2.44\times0.83$ & 29 & \tablenotemark{h} & 3\\
~~~~~~~~~~~~~$^1\nu_{15}$ & $25_{0,25}-24_{0,24}$ & 231,235.325 (75) & 635 & 363.3 & $2.44\times0.83$ & 29 & \tablenotemark{h} & 3\\
$^{13}$CS & $5-4$ & 231,220.996 (200) & 33 & 19.3 & $2.44\times0.83$ & 29 & \tablenotemark{h} & 3\\
\etcn & $26_{1,25}-25_{1,24}$ & 231,310.420 (50) & 153 & 383.0 & $2.44\times0.83$ & 29 & \tablenotemark{h} & 3\\
\enddata
\tablenotetext{a}{When compared with the results from \citet{sutton87}.}
\tablenotetext{b}{These \dme\ lines are blended.}
\tablenotetext{c}{Calculations indicate that little or no flux is missing, but since the features are unresolved in our observations and source sizes cannot be determined the values are upper limits.}
\tablenotetext{d}{These \acetone\ lines are blended.}
\tablenotetext{e}{Not detected by \citet{sutton87}.}
\tablenotetext{f}{The full line was not detected in our observations, so no comparison could be made.}
\tablenotetext{g}{This line was not detected in our observations, so no comparison could be made.}
\tablenotetext{h}{Since this line was spectrally unresolved, no comparison can be made.}
\tablerefs{(1) \citet{oest99}; (2) \citet{gron98}; (3) \citet{jpl}; (4) \citet{gron02}; (5) \citet{xu97}; (6) \citet{lovas04}}
\label{tab:mols}
\end{deluxetable}

\clearpage
\begin{deluxetable}{lccccccrccr}
\tablecolumns{11}
\tablewidth{0pt}
\tabletypesize{\scriptsize}
\tablecaption{Fits and Column Densities of Observed Lines}
\tablehead{\colhead{} &\colhead{} &\colhead{} &\colhead{} &\colhead{} &\colhead{} &\multicolumn{2}{c}{Low T\tablenotemark{b}} &\colhead{}&\multicolumn{2}{c}{High T\tablenotemark{c}}\\
\cline{7-8}\cline{10-11}\colhead{} & \colhead{} & \colhead{} & \colhead{$I_0$\tablenotemark{a}} & \colhead{$v_{\rm LSR}$\tablenotemark{a}} & \colhead{${\Delta}v$\tablenotemark{a}} & \colhead{} & \colhead{$N_T$} & \colhead{} & \colhead{} & \colhead{$N_T$}\\
\colhead{Molecule} & \colhead{Source} & \colhead{Fig.} & \colhead{(Jy bm$^{-1}$)} & \colhead{(km s$^{-1}$)} & \colhead{(km s$^{-1}$)} & \colhead{$\tau$} & \colhead{(cm$^{-2}$)} & \colhead{} & \colhead{$\tau$} & \colhead{(cm$^{-2}$)}}
\startdata
\etcn & IRc7 & \ref{fig:etcn}b & 2.5(2) & 4.9(1) & 2.8(2) & 0.36(4) & 3.2(4)$\times10^{16}$ & & 0.13(1) & 1.9(3)$\times10^{17}$\\
 & IRc7 & \ref{fig:etcn}b & 0.8(2) & 3.9(8) & 11.6(11) & 0.10(3) & 3.7(10)$\times10^{16}$ & & 0.04(1) & 2.4(6)$\times10^{17}$\\
 & Hot Core & \ref{fig:etcn}c & 3.9(1) &5.1(2) &6.3(1) & 0.74(4) & 1.5(1)$\times10^{17}$ & & 0.21(1) & 7.0(3)$\times10^{17}$\\
\dme & IRc6 & \ref{fig:dme}b & 1.8(4)\tablenotemark{d} &7.1(1) &1.5(4) & 0.63(20) & 9.0(3)$\times10^{17}$ & & 0.35(10) & 1.3(5)$\times10^{18}$\\
 & IRc6 & \ref{fig:dme}b & 2.8(3) &9.1(4) &6.0(6) & 1.29(28) & 7.4(13)$\times10^{18}$ & & 0.62(9) & 9.3(14)$\times10^{18}$\\
 & IRc5 & \ref{fig:dme}c & 3.4(4) &7.6(1) & 1.1(2)& 1.77(65) & 1.9(6)$\times10^{18}$ & & 0.78(15) & 2.1(5)$\times10^{18}$\\
 & IRc5 & \ref{fig:dme}c & 0.9(3) &8.1(6) & 2.5(14)& 0.25(10) & 6.0(40)$\times10^{17}$ & & 0.16(6) & 9.6(64)$\times10^{17}$\\
\mef\tablenotemark{e} & IRc6 & \ref{fig:mef}b & 0.5(1) & 7.7(4) & 3.3(8) & 0.13(3) & 3.1(10)$\times10^{16}$ & & 0.08(2) & 4.6(15)$\times10^{16}$\\
 & IRc5 & \ref{fig:mef}c & 3.7(3) & 7.5(0) & 1.3(1) & 2.11(7) & 2.0(5)$\times10^{17}$ & & 0.83(11) & 1.8(2)$\times10^{17}$\\
 & IRc5 & \ref{fig:mef}c & 2.4(4) & 9.3(0) & 0.9(1) & 0.84(23) & 3.6(8)$\times10^{17}$ & & 0.45(10) & 4.4(9)$\times10^{17}$\\
\mtoh & IRc5 & \ref{fig:mtoh}b & 4.1(5) & 7.8(2) & 6.6(6) & 4.37(1030) & 1.5(33)$\times10^{18}$ & & 1.00(22) & 9.2(17)$\times10^{17}$\\
 & IRc5 & \ref{fig:mtoh}b & -4.1(3) & 8.6(1) & 1.4(3) & \tablenotemark{f} & & & \tablenotemark{f} & \\
 & IRc6 & \ref{fig:mtoh}c & 2.3(4) & 7.1(2) & 1.4(7) & 0.80(22) & 5.9(32)$\times10^{16}$ & & 0.44(10) & 8.6(46)$\times10^{16}$\\
 & IRc6 & \ref{fig:mtoh}c & 5.0(4) & 8.7(2) & 0.6(3) & \tablenotemark{g} &  & & 1.47(29) & 1.2(6)$\times10^{17}$\\
 & IRc6 & \ref{fig:mtoh}c & 1.5(3) & 10.3(2) & 1.3(4) &  0.45(12) & 3.1(11)$\times10^{16}$ & & 0.26(6) & 4.8(18)$\times10^{16}$\\
 & Hot Core & \ref{fig:mtoh}d & 2.8(2) & 5.6(2) & 5.5(4) & 0.42(4) & 7.4(8)$\times10^{17}$ & & 0.13(1) & 4.7(5)$\times10^{18}$\\
 & Hot Core & \ref{fig:mtoh}d & -1.8(4) & 7.4(1) & 1.1(3) & \tablenotemark{f} & & & \tablenotemark{f} & \\
 & IRc7 & \ref{fig:mtoh}e & 1.3(3) & 4.9(8) & 6.9(16) & 0.16(4) & 3.5(11)$\times10^{17}$ & & 0.06(1) & 2.5(9)$\times10^{18}$\\
 & IRc7 & \ref{fig:mtoh}e & -1.2(6) & 7.0(3) & 1.2(9) & \tablenotemark{f} &  & & \tablenotemark{f} &\\
\acetone & IRc7 & \ref{fig:ace}b & 1.3(2)\tablenotemark{h} & 8.0(2)& 2.4(4) & 0.37(6) & 3.2(7)$\times10^{17}$ & & 0.23(3) & 5.5(12)$\times10^{17}$\\
 & IRc7 & \ref{fig:ace}b & 0.7(2)\tablenotemark{h} & 5.3(4) & 2.4(10) & 0.18(5) & 1.6(8)$\times10^{17}$ & & 0.12(3) & 2.8(14)$\times10^{17}$\\
 & Hot Core-SW & \ref{fig:ace}c & 1.7(2)\tablenotemark{h} & 7.7(2) & 2.1(4) & 0.75(15) & 5.7(13)$\times10^{17}$ & & 0.38(6) & 8.1(18)$\times10^{17}$\\
 & Hot Core-SW & \ref{fig:ace}c & 0.4(2)\tablenotemark{h} & 5.3(8) & 2.5(6) & 0.14(7) & 1.3(7)$\times10^{17}$ & & 0.08(4) & 2.1(11)$\times10^{17}$\\
\enddata
\tablenotetext{a}{All uncertainties listed are 2$\sigma$, an uncertainty of 0 means the there are not enough significant figures to display its true value.}
\tablenotetext{b}{Low temperature for hot core/IRc7 is 150K, for all other sources is 100 K}
\tablenotetext{c}{High temperature for hot core/IRc7 is 500K, for all other sources is 150 K}
\tablenotetext{d}{$I_0$ for blend of all torsional states.}
\tablenotetext{e}{Quantities are for both A and E torsional states, as they are assumed to have equal populations.}
\tablenotetext{f}{Opacity and column density not calculated for absorption components.}
\tablenotetext{g}{Opacity calculation gave negative values, likely due to too low an assumed rotation temperature.}
\tablenotetext{h}{$I_0$ for EE torsional state only; for other states as follows: $I_0$(AA) = 0.5$I_0$(EE), $I_0$(EA) = $I_0$(AE) = 0.25$I_0$(EE).}
\label{tab:fits}
\end{deluxetable}

\clearpage

\clearpage

\clearpage

\clearpage

\clearpage

\clearpage

\clearpage

\clearpage

\clearpage

\clearpage

\clearpage
\end{document}